# Net Shape 3D Printed NdFeB Permanent Magnet

J. Jaćimović[1*], F. Binda[1], L. G. Herrmann[1], F. Greuter[1], J. Genta[1], M. Calvo[1], T. Tomse[2,3], R. A. Simon[1]

[1]*ABB Corporate Research Center, Baden-Daettwil, Switzerland*
[2]*Department for Nanostructured Materials, Jožef Stefan Institute, 1000 Ljubljana, Slovenia*
[3]*Jožef Stefan International Postgraduate School, 1000 Ljubljana, Slovenia*

For two decades, NdFeB based magnets have been a critical component in a range of electrical devices engaged in energy production and conversion. The magnet shape and the internal microstructure of the selected NdFeB grade govern their efficiency and size. However, stricter requirements on device efficiency call for better performing magnets preferably with novel functionality not achievable today. Here we use 3D metal printing by Selective Laser Melting to fabricate dense net shape permanent magnets based on NdFeB that exhibit high magnetic performance. Evidence is provided that the internal microstructure, not achievable by traditional manufacturing means, is the origin of the solid magnetic properties. The freedom in magnet body shape and size that ranges from the millimeter to tens of centimeter scale opens up a design freedom that could be a catalyzer for the next generation of electrical devices.

## Introduction

NdFeB magnets exhibit remarkable performances that are superior to previously discovered hard magnets, like Alnico and hard ferrites [1] (magnetic remanence $B_r$ higher than 1 T, coercivity $H_c$ above 1000 kA/m, and figure of merit $(BH)_{max}$ larger than 200 kJ/m$^3$). The other magnets need to have at least a five times larger volume than NdFeB to achieve the same magnetic performances. Therefore, this magnet emerges as one of the key components that could enable further miniaturization and increase in efficiency of electrical devices for energy production and conversion. Mass market consumer electronic products like cell phones, hard drives, cameras and sensors broadly rely on NdFeB for data storage and sensing [2, 3] already today. The same magnet figures as the best candidate to be integrated in high efficient motors in electrical vehicles and robots, aiming to reduce their weight and increase robustness [4] [5]. In offshore wind turbine generators, where light and efficient components are an obligation, NdFeB permanent magnets also emerge as the most promising solution [6]. Finally, this material is exploited in medical applications like Magnetic Resonance Imaging (MRI), where it is considered as the most serious alternative to replace expensive superconducting magnets used nowadays [7].

To respond to the aforementioned emergent demands and to unlock the full potential of the technology, there has been an ever-increasing industrial and scientific effort to design a new generation of NdFeB magnets. Two main goals were identified: first, complex net shape magnets and second, superior performance stability at elevated temperatures up to 200 ºC [8].

Concerning the former goal, Toyota has recently demonstrated an increase of the motor output power by 50 %, if each two neighbouring permanent magnets in the rotor are arranged in a so called "V" shape geometry compared to the traditionally used flat – shape arrangement [9]. Even additional gain can be achieved, if these "V" magnets would be replaced by a single magnetic body of a curved shape [10]. However, sintered magnets used nowadays are difficult to be fabricated with a complex net geometry as the green magnetic bodies lose up to 25 volume % during the sintering procedure and they are rather brittle and challenging to be machined after the sintering [11]. Number of alternative methods for complex shape magnets have been tested to date. The most promising are bonding magnetic composites and Spark Plasma Sintering

(SPS). The former process offers substantial design freedom, but the magnet is limited in operation due to thermal and mechanical stability of the polymer. Additionally, the magnetic body suffers from low magnet density (typical between 60 and 80 volume %), that reduces the overall magnet efficiency [12, 13]. The SPS method produces the magnets free of polymer matrix but, the technology is limited in geometry variations [14].

The second goal concerns sensors and other electrical devices that operate above 100 ºC, where the coercivity in NdFeB magnet decreases rapidly with the rise of temperature, risking loss of the material functionality [2, 12]. To date, two encouraging avenues have been identified to be most suitable to overcome the issue. First, doping with the heavy rare earth element Dy (5-10 weight %) was successful in rising coercivity above 2000 kA/m but, it decreased remanence at room temperature below 1 T [15], and at the same time increased the magnet price by *circa* 30 % (Dy is 10 times more expensive than Nd). Alternatively, Ramesh *et al.* proposed to reduce the grain size in the sintered magnets, showing that the coercivity increases as the grain size decreases, reaching its maximum at 250 nm, when the grain size is comparable to the magnetic domain size [16]. However, all commercial sintered magnets have grain sizes at least one order of magnitude above this value (higher than 10 µm) [17], because of the oxygen sensitivity and grain growth during the sintering.

## Experimental Section

The commercial *spherical powder* of the chemical composition $Nd_{7.5}Pr_{0.7}Zr_{2.6}Ti_{2.5}Co_{2.5}Fe_{75}B_{8.8}$, known in the market as MQP-S and supplied by Magnequench Corporation, is used in the study. In terms of weight percent, it contains 19 % of Rare Earth elements Nd and Pr. The powder size distribution showed that $d_{50}$ is 38 µm. The flow rate of 3.33 g/cm$_3$ is comparable to other powders used for 3D printing like CoCr and stainless steel. The measured tap density is 4.67 g/cm$^3$ corresponding to 61.4 % of the full density.

A commercial *metal 3D Printer* (Model Realizer SLM 50) was used to produce three dimensional magnetic samples. Before the printing process starts, a design of the required object was created in a CAD software and sliced in layers. The latter is an important step as it determines the thickness of each powder layer (LT) that will be deposited on the printing platform and be molten by the laser beam. The quality of the printed material is determined by five laser parameters: Laser Power (LP), Laser Focus (LF), Point Distance (PD), Exposure Time (ET) and Hatching Distance (HD). A pulsed Nd:YAG-Laser with a maximum energy output of 120 W LP is used to melt the powder. Its intensity, LP, is controlled by setting the current from 500 to 3000 mA. LF describes how the laser light is concentrated to the powder and the spot size diameter at the sample surface varies from 15 µm to 30 µm. The model of 3D printer we used has a discrete motion of the laser that is controlled by a scanner. The movement of the laser focus from one point to the next is defined by the PD that can take values from 1 to 50 µm. ET that ranges from 10 to 300 µs, describes the time duration of the spot exposed by the laser beam. Defining these 4 parameters is necessary to print a "one dimensional" track. To continue printing in 2 dimensions the laser moves aside the track and starts melting the neighboring raw powder. This move is named hatching distance HD. Finally, the printing procedure is executed under argon gas with an oxygen level below 500 ppm.

*Spark Plasma Sintering* was performed using a commercial set-up, (Dr. Sinter Spark Plasma System, Fuji Electronic Industrial Co. Ltd). The powder was placed in a cylindrical graphite mould (16 mm in diameter), pressed with 50 MPa and, positioned inside the sintering machine. Sintering is done by sending high frequency electrical current pulses (pulse length 0.1s, current amplitude of 1100 A) through the sample. The sintering/holding time was 1 min. The processing temperature was monitored using a pyrometer. A temperature of 800°C was found to be optimal for the MQP-S powder. During sintering, the pressure inside the chamber was maintained at 0,3 mbar.

*Magnetic measurements* were done with a Pulsed Field Magnetometer (PFM), HyMPulse from Metis. A capacitor bank was used to generate current pulses leading to magnetic sinusoidal pulses up to 7 T with a duration of several milliseconds. Two pulses are used to complete the measurement. The first pulse is used to magnetize the material, while the purpose of the second is to reverse the magnetic orientation in the sample. Pick-up coils detect the time dependence of the magnetization reversal from which the magnetic characteristics of the examined material can be calculated.

*Scanning electron microscope* (SEM) was used to analyse the surface morphology of the prepared materials. The accelerating voltage was tuned between 2 and 10 kV to obtain high quality images. To investigate and map the chemical composition of the sample surface Energy-Dispersive X-ray Spectroscopy (EDX) was used.

*X-ray diffraction* (XRD) was used to characterize the crystal structure of the material. The measurements were carried out at ambient conditions) using the $\theta$–$2\theta$ technique with Cu $K\alpha$ radiation where the angle $2\theta$ was changed between 10 and 90° in steps of 0.02°.

## Results and Discussion

In this work, we show for the first time that the unconventional process, Selective Laser Melting (SLM) of metals can be used to 3D print dense, free shape magnets of very fine microstructure [18], with stable magnetic performances up to elevated temperatures. To prepare high quality objects, it must be ensured that the powder used for the printing has spherical morphology to enable powder flowability necessary for homogenous and dense deposition of a powder bed on a printing platform [19]. Only such powders can lead to stably printed objects without severe crack and pore formation [20]. The only available NdFeB material with such property is a commercial powder (known as MQP-S) of chemical composition $Nd_{7.5}Pr_{0.7}Zr_{2.6}Ti_{2.5}Co_{2.5}Fe_{75}B_{8.8}$, developed for polymer bonded magnets. In terms of the weight percent, it contains 19 % of Rare Earth elements (Nd and Pr), that is 7 % less than in nominal $Nd_2Fe_{14}B$, and even 12 % lower than in commercial sintered magnets [21] used in highly efficient electrical devices phase (details of the powder characteristics are given in the experimental section). The magnetic samples were printed on a steel platform connected with a larger copper piece. Such architecture serves as a heat sink to evacuate the laser-produced heat so that the platform is not overheated. To 3D print reliably fully dense functional permanent magnets of the very fragile NdFeB alloy, it is of upmost importance to discover the appropriate printing conditions. In our case this was done by a systematic variation of six independent printing parameters: laser power (LP), laser focus (LF), point distance of the laser movement (PD), time that powder is radiated by the laser (ET), thickness of deposited fresh powder layer (LT) and horizontal movement of the laser – Hatching Distance (HD) **Figure 1a**. The unique combination of LP=1700 mA, LF=30 µm, PD=30 µm, ET=110 µs, LT=20 µm and HD=100 µm ensured very focused micrometre size melting volume, that can cool down very fast with a rate in the range of typically $10^4 – 10^6$ °C/s [22]. The printed $Nd_2Fe_{14}B$ phase has a grain size of only 1 µm, that is one order of magnitude lower than in sintered magnets. This leads to very good magnetic properties of the as printed samples without additional post heating treatment: $H_c$= 695 kA/m,

$B_r = 0.59$ T, $(BH)_{max}=45$ kJ/m³, which is the maximum with the used powder formulation, **Figure 1b** (over dozens of measured samples, 3 % variations in magnetic properties were noted). The density of the produced sample was 92 %. On 5x5x5 mm³ cubes 50 µm long microcracks were detected. As they are not severe, the cracks did not prevent the scale up of the printed specimens. In fact, the discovered printing conditions enabled printing stable magnets as large as 3000 mm³. To demonstrate the full power of 3D printing (digital manufacturing), very complex shapes, with channels inside the object (e.g. potential cooling paths), were successfully printed **Figure 1c**.

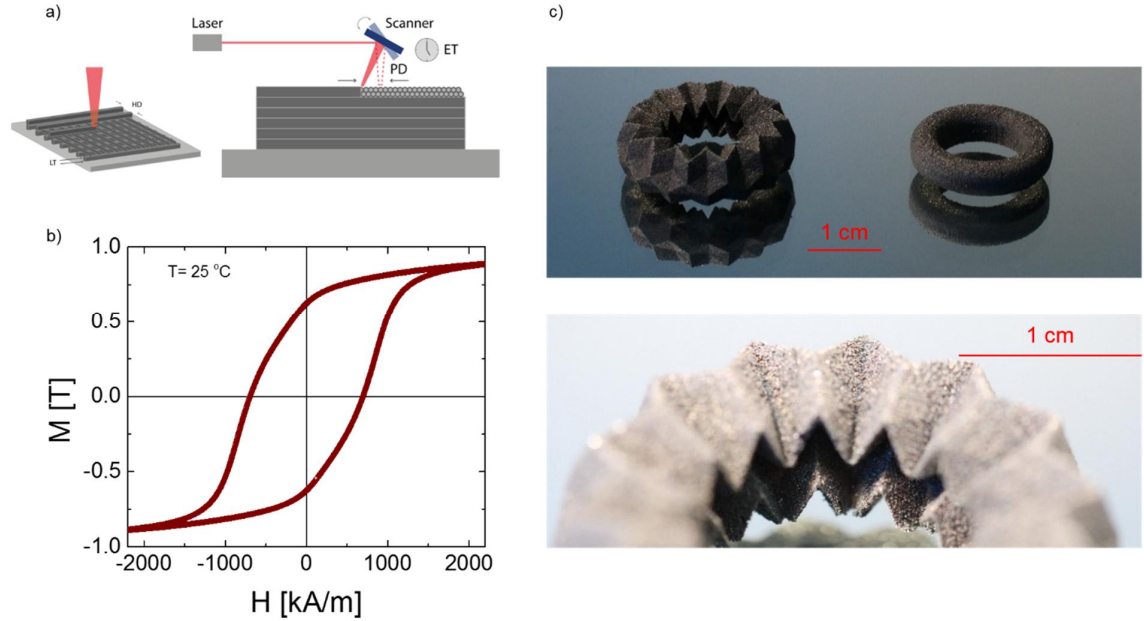

*Figure 1: a) Schematic illustration of the 3D printing method of magnet fabrication. b) Full magnetic hysteresis loop for 3D printed sample with $H_c$=695 kA/m and $B_r$= 0.62 T, at room temperature. c) The photograph of printed magnets of various shapes. The left object has a channel that adds a novel functionality to the hard magnets (see zoom on bottom right part of figure).*

**Figure 2a)** shows how the magnetic figure of merit depends on the energy density ($E = LP/(LF * LT * HD * PD/ET)$) of the laser supplied to the powder during the printing procedure. The general trend is that the magnetic properties improve as *E* increases. However, since the energy density takes into account six independent printing parameters, a more detailed analysis of the specific parameter contribution is needed to fathom the correlation between them and magnetic properties. As **Figure 2** depicts for the same energy density, different magnetic properties can be achieved when tuning one specific laser parameter. Two parameters are particularly important, namely laser velocity ($v = PD/ET$) and layer thickness (LT). For illustration, we define three different regions in Figure 2. In zone I, for large energy density and low laser velocity (v=0.025 m/s), the magnetic properties of the printed samples are very weak and vary around Hc=20 kA/m, Br=0.08 T, and $(BH)_{max}$=0.1 kJ/m³. In area II, the laser velocity is one order of magnitude higher than in the previous region leading to an increase of $(BH)_{max}$ by two orders of magnitude. The maximum $(BH)_{max}$ of 40 kJ/m³ was achieved for a laser velocity of v=0.36 m/s. Finally, region III depicts that layer thickness, LT, is another effective parameter for the fine-tuning of the magnetic properties, keeping constant the optimum laser speed previously identified. As LT decreases from 100 to 20 µm, $(BH)_{max}$ rises from 15 to 45 kJ/m³. These results suggest that the process parameters define the manner in which the molten powder solidifies, and hence determine the magnetic characteristics of the printed objects accordingly. It is worth mentioning that for every set of parameters more than 5 samples were measured, and variation in magnetic characteristics from sample to sample is 2 – 3 %. To further investigate this hypothesis, we analysed the microstructural details and the shape of different phases in the samples with lowest and highest magnetic properties, see **Figure 2 c-d)**.

**Figure 2c** corresponds to a cross section of a specimen from region I in Figure 2a, with Hc=10 kA/m, Br= 0.05 T, and (BH)$_{max}$ of only 0.1 kJ/m$^3$. Here, the laser takes more than a second to completely move away from the liquid pool, which is thus rather deep, 100 µm, and the solidification process is consequently slow. By means of EDX mapping and XRD analysis, three different material phases on the investigated cross section are clearly identified. The small round like white regions of a few hundreds of nanometres extension correspond to neodymium oxide. Interaction between the residual oxygen in the printing chamber (below 500 ppm) and very reactive rare earth metals may be the origin of the oxide formation. This phase is not magnetic but still perturbs the coercivity of the sample via reversal of the magnetic domains [23]. It also influences the remanence, which is proportional to the volume percent of the magnetic phase in the specimen. Further, more two phases were recognised in the cross section; Fe and Nd$_2$Fe$_{14}$B, in dark and light grey colours, respectively. The former is typically a few microns long (the maximum size is 30 µm) and has dendritic like structure, whereas the latter has irregular shape of a micro meter extension size. The development and coexistence of these two phases could be followed on the equilibrium Nd-Fe-B phase diagram, **Figure 2b**. The low laser speed justifies the assumption that the solidification process after printing is slow, so that conditions close to the thermodynamic equilibrium are established. The MQP-S powder used with only 7 atomic percent of rare earth elements is positioned on the left of the Nd$_2$Fe$_{14}$B line. On this side, there is no temperature interval where the wanted hard magnetic phase $\Phi_1$ grows as the single phase. On the contrary, the iron is dominant. Fe starts to solidify directly from the liquid, and grows solely above the peritectic temperature of 1181 ºC, when Nd$_2$Fe$_{14}$B grows. These two phases solidify and remain stable down to room temperature (it is important to state that the existence of other phases like Iron-Boride or Nd$_x$Fe$_y$B in the sample microstructure is not excluded however, they are in very small amount and hard to detect and its role goes beyond the scope of this study). The presence of iron in the matrix of Nd$_2$Fe$_{14}$B has two negative effects on the magnetic properties of the printed samples. First, it reduces the overall volume percent of the hard magnetic phase and consequently leads to a reduced remanence. Second, even more severe that total coercivity is reduced since the soft magnetic iron can act as the centre of reversed magnetic domain in the hard magnetic phase [24].

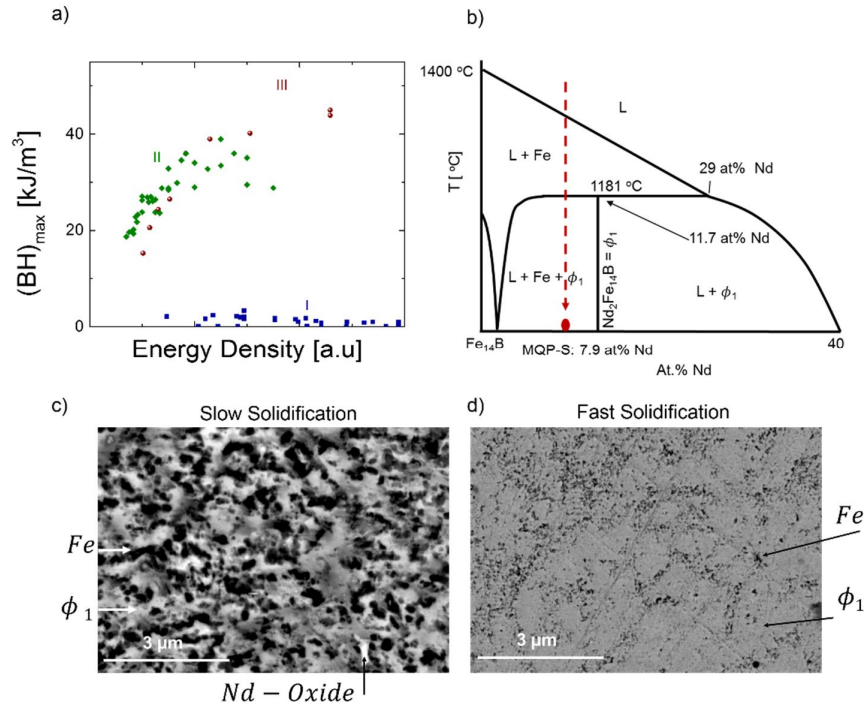

*Figure 2: **a**) Laser energy density dependence of magnetic figure of merit $(BH)_{max}$. **b**) Schematic view on the $Fe_{14}B$-Nd phase diagram, where $\Phi_1$ is desired hard magnetic phase, **c**) SEM picture of the printed sample from the zone I in Figure 2a) with slow solidification. **d**) SEM picture of the printed sample from the zone III in Figure 2a) with fast solidification.*

**Figure 2d** shows the cross section SEM picture of a sample from region III in **Figure 2a**, with $H_c$= 695 kA/m, $B_r$ = 0.59 T, and $(BH)_{max}$=45 kJ/m$^3$. The time the powder spot was exposed to the laser during the printing was only 100 µs. The proper laser power and low thickness of the deposited powder layer (20 µm) lead to a rather thin liquid pool (20-30 µm) that was in direct contact with previously solidified material. Such circumstances gave rise to a rapid solidification of the liquid producing non-equilibrium conditions [25]. To stabilize the growth of the target magnetic phase $\Phi_1$, powder additives like Co, Zr and Ti play a significant role [26]; acting as catalyser for the nucleation of $Nd_2Fe_{14}B$ and further influence the kinetics and thermodynamics of the solidification process. The high cooling rate and the particular composition of the MQP – S alloy resulted in stabilisation of the $\Phi_1$ magnetic phase was (light grey regions) even though the rare earth concertation was 10 atomic percent; i.e. lower than the minimum required by the peritectic (17 atomic percent) [27]. Small iron segregation (dark grey regions) is still observed, but it is for two orders of magnitude smaller than in the previous slow cooling scenario.

We now turn to compare the magnetic performances of 3D printed, injection molded, and Spark Plasma Sintered magnets, produced from the same MQP-S powder. Direct comparisons with a traditionally sintered magnet are not done, as this powder is not suited for this process. The room temperature hysteresis curves in the second quadrant for the three samples are presented in Figure 3. The higher squareness of the hysteresis curve for the injection molded magnet could be explained by the absence of free iron in the starting powder, whereas traces of it are detected in 3D printed samples (as discussed above). Magnetic features such as coercivity and remanence are highest in 3D printed magnets. Hc is 5 and 15 % higher than in the injection molded and SPS magnets, respectively. The significantly lower coercivity in SPS magnet is explained by the grain growth during the sintering [28]. In terms of remanence, the 3D printed magnet outperforms the polymer bonded magnet by 15 and the SPS-sintered magnet by 18 %. This is primarily the consequence of the impressive 92 % in volume density of the printed magnet, while in the injection molded and SPS magnets we observe densities of 65 and 70-75 %, respectively.

The temperature dependence of $(BH)_{max}$ of these three samples is presented in **Figure 3b**. At room temperature 3D printed and injection molded magnets have similar figures of merit which is 50 % larger than for the SPS sample. With rise of temperature, $(BH)_{max}$ decreases in all three samples, as expected for this alloy formulation. However, the injection molded magnet (prepared with mostly used polymer Polyamide 12 – PA 12) is limited in typical applications to only 120 – 140 ºC by thermal and mechanical properties of the polymer matrix, whereas 3D printing offers a much broader operational temperature range that makes these magnets suitable for highly demandable applications. Additionally, 3D printing opens new design possibilities like fabrication of cooling channels within the magnet, or optimized mechanical and magnetic flux design in integrating sensors. Internal cooling of the magnet in power applications could be used to prevent overheating and improve the efficiency of electrical devices even further. In the long term, such technology could help to reduce or, even completely eliminate the use of expensive heavy rare earth elements.

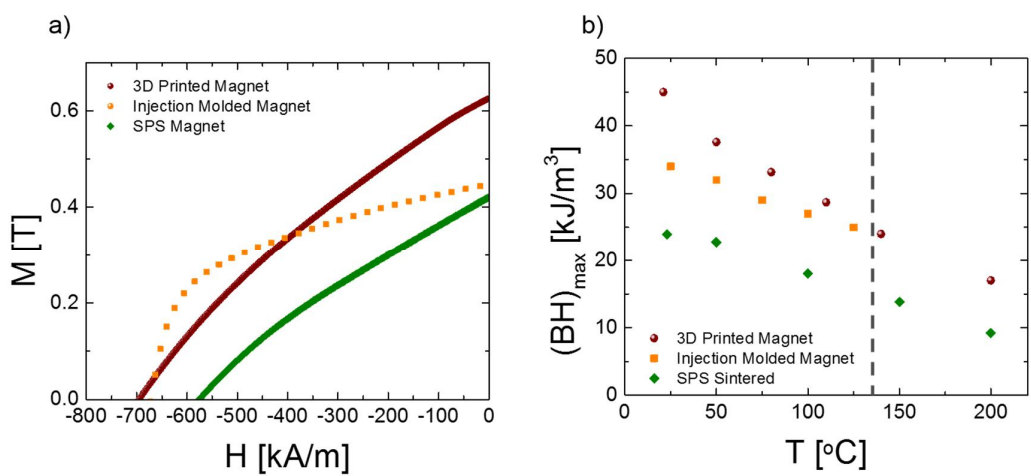

*Figure 3: **a**) Magnetization as the function of demagnetizing field for three different samples, all based on the same MQP – S powder: 3D printed, Injected molded magnet with most used PA 12 polymer, and SPS sintered. **b**) Temperature dependence of the $(BH)_{max}$ for the same three samples. The data for the injection molded magnet are taken from the reference [29].*

## Conclusion

In summary, we have realized for the first time 3D printed hard magnets by Selective Laser Melting of metals. Digital manufacturing of the advanced Nd-Fe-B magnet is a particular challenge due to pronounced peritectic in the phase diagram. Fine-tuning of the laser parameters like laser speed, focus and power, together with the control of the thickness of the deposited powder layer, allowed rapid production of the dense net shape hard magnets with the grain size of ~1µm, a property not achievable with standard manufacturing processes. The impressive magnetic characteristics of the printed samples outperform injection molded and SPS sintered magnets composed of the same powder. The high temperature magnet stability and "the net shape freedom" could enable integration, miniaturization and efficiency increases of electronic and electrical devices. Our findings may open novel routes for further technology progress, like dedicated generators producing "green" electricity, or highly efficient electromotors for transportation and robotics. Our results pave the way for development of next generation of powerful 3D printed magnets, as printing the NdFeB powder with higher amount of rare earth elements (above the peritectic composition) could double the current coercivity and remanence.

The new freedom in designing the shape and the magnetic flux pattern in high performance permanent magnet is expected to establish a new paradigm in energy related technologies.

## Acknowledgment

We thank Andrijana Drobnjak for preparing the graphic illustrations, then to Cornelia Lang and Fabian Haag (ETHZ) for acquiring SEM images. Furthermore, we would like to thank Jens Rocks, Darren Tremelling, Elio Perigo and Daniel Chartouni for fruitful discussions and support as well as Daniel Kearney for proof-reading of the manuscript.

* jacim.jacimovic@ch.abb.com